\documentclass[slac_one]{revtex4}
\usepackage{graphicx}
\usepackage{fancyhdr}
\begin{document}

\title{Semileptonic B decays at Belle} 

%

\author{W. Dungel}
\affiliation{Institute of High Energy Physics, Austrian Academy of Sciences}

\begin{abstract}
We present three analyses based on data samples collected by the Belle detector at the KEKB $e^+e^-$ asymmetric energy collider.
The first one contains the measurement of the branching fractions and $q^2$ distributions of exclusive charmless semileptonic B decays, which are used to extract the Cabibbo-Kobayashi-Maskawa matrix element $|V_{ub}|$. The sample of events is obtained by fully reconstructing one of the B mesons in selected hadronic modes.
The second analysis determines the branching fraction and the form factors $\rho^2$, $R_1$ and $R_2$ for the exclusive decay $B^0 \rightarrow D^{*-} \ell^+ \nu$. These measurements are used to extract the Cabibbo-Kobayashi-Maskawa matrix element $|V_{cb}|$.
The third analysis investigates $B \rightarrow D^{**} \ell \nu$ decays and determines branching ratios for the four modes $D_{0}^{*}$, $D_{1}$, $D_{1}^{'}$ and $D_{2}^{*}$ for both charged and neutral B decays.

\end{abstract}

\maketitle

\thispagestyle{fancy}


\section{KEKB and the Belle detector} 

The following analyses were performed using data samples collected at the $\mathcal{Y}(4S)$ resonance with the Belle detector operating at the KEKB asymmetric-energy $e^+ e^-$ collider. The Belle detector is a large-solid-angle magnetic spectrometer that consists of a silicon vertex detector (SVD), a 50-layer central drift chamber (CDC), an array of aerogel Cherenkov counters (ACC), a barrel-like arrangement of time-of-flight scintillation counters (TOF) and an electromagnetic calorimeter comprised of CsI(Tl) crystals (ECL) located inside a superconducting solenoid coil that provides a 1.5 T magnetic field. An iron flux-return located outside the coil is instrumented to detect $K_L^0$ mesons and to identify muons (KLM). The detector is described in detail in Ref.~\cite{BelleDetector}.

\section{Analyses of semileptonic B decays}


\subsection{Exclusive $B\rightarrow X_u \ell \nu$ decays}

One of the two $B$ mesons is reconstructed in a hadronic decay cascade (``full reconstruction''), yielding the so-called tag momentum, $p_{\mathrm{tag}}$. 
The direction of the signal $B$ in the c.m. frame is assumed to be opposite of the tag meson. The energy is set to equal the beam energy, $E_{\mathrm{beam}}$. The magnitude of the four momentum is defined via the nominal $B$ mass, $m_{B}$. The remaining event is then used to reconstruct both a light lepton $\ell$, either $e$ or $\mu$, and a meson $X$, which is one of the five light mesons $\pi^{\pm}$, $\pi^0$, $\rho^\pm$, $\rho^0$ or $\omega$.

\begin{table}[hb]
\begin{center}
\caption{Partial branching fractions in 3 bins of $q^2$, obtained from the fit. These are summed to give the full branching fraction quoted in the ``Sum'' column. Errors are statistical and systematic. The numbers are premliniary.}
		\label{tab:KevinResults}
    \begin{tabular}{|l|@{\extracolsep{.25cm}}c| c| c|c|}
      \hline 
      \phantom{o} & \multicolumn{3}{c|}{$\Delta \mathcal{B} [10^{-4}]$} & $\mathcal{B} [10^{-4}]$\\
      \hline
      \phantom{o} & $0<q^2<8$ & $8<q^2<16$ & $q^2<16$ & Sum\\
      \phantom{o} & $[GeV^2]$ & $[GeV^2]$  & $[GeV^2]$& \\
      \hline
      $B \rightarrow \pi^{+} \ell \nu$ & $0.43 \pm 0.11 \pm 0.02$ & $0.42 \pm 0.11 \pm 0.02$
       & $0.26 \pm 0.08 \pm 0.01$ & $1.12 \pm 0.18 \pm 0.05$ \\
      $B \rightarrow \pi^{0} \ell \nu$ & $0.26 \pm 0.09 \pm 0.01$ & $0.17 \pm 0.05 \pm 0.01$
       & $0.22 \pm 0.06 \pm 0.01$ & $0.66 \pm 0.12 \pm 0.03$ \\
      $B \rightarrow \rho^{+} \ell \nu$ & $0.74 \pm 0.29 \pm 0.04$ & $1.01 \pm 0.28 \pm 0.05$
       & $0.81 \pm 0.21 \pm 0.04$ & $2.56 \pm 0.46 \pm 0.12$ \\
      $B \rightarrow \rho^{0} \ell \nu$ & $0.72 \pm 0.15 \pm 0.03$ & $0.70 \pm 0.13 \pm 0.03$
       & $0.39 \pm 0.11 \pm 0.02$ & $1.80 \pm 0.23 \pm 0.07$ 
      \\ $B \rightarrow \omega \ell \nu$ & $0.23 \pm 0.17 \pm 0.01$ & $0.64 \pm 0.21 \pm 0.03$
       & $0.32 \pm 0.17 \pm 0.01$ & $1.19 \pm 0.32 \pm 0.05$ 
      \\ \hline
    \end{tabular}
\end{center}
\end{table}

\begin{figure}[h]
  \caption{The missing mass squared distributions for the five exclusive reconstruction modes. The sum of the individual fits in three bins of $q^2$ is shown.}
  \label{fig:KevinMissingMass}
  \begin{center}
    \includegraphics[width=0.20\columnwidth]{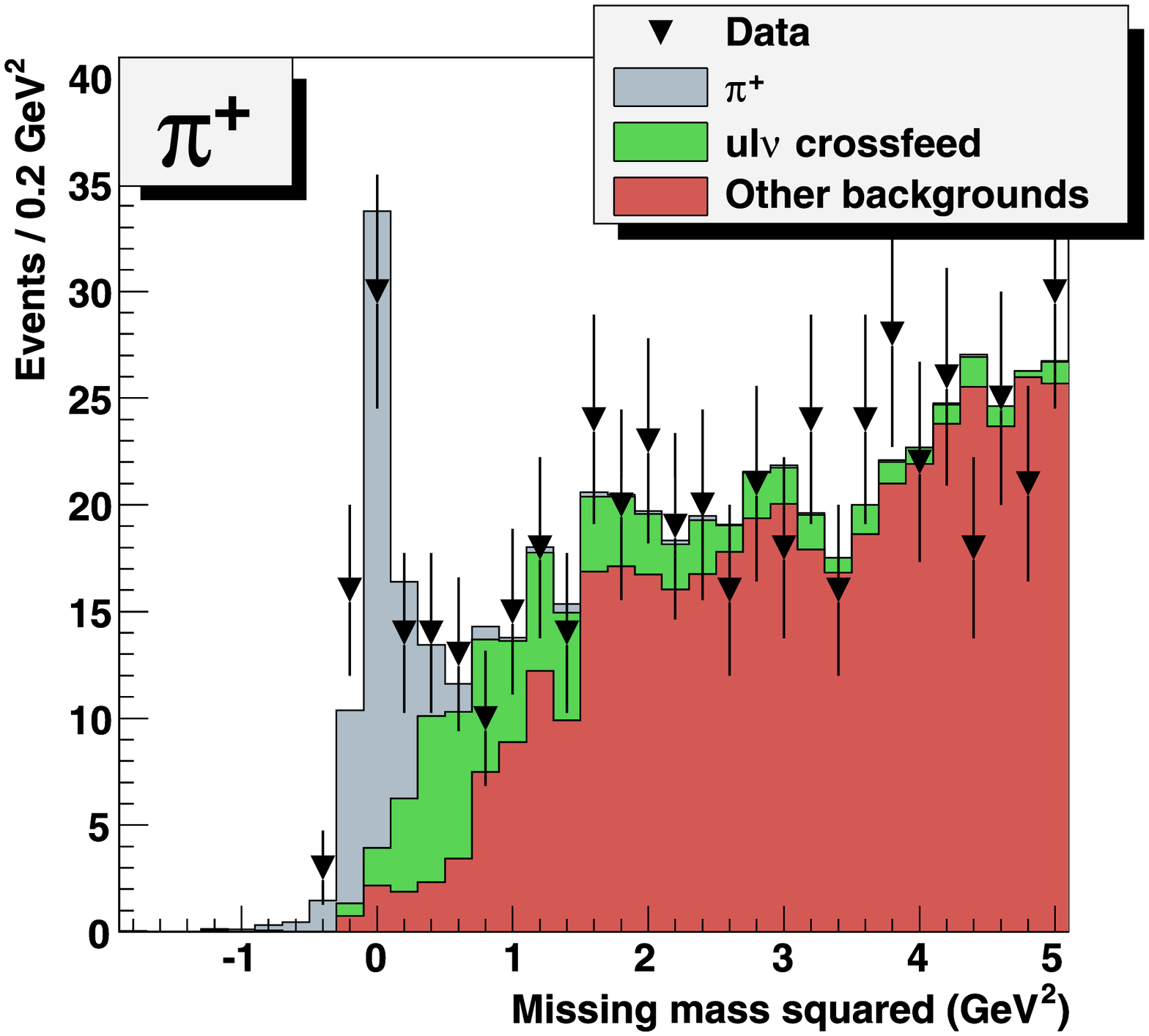}
    \includegraphics[width=0.20\columnwidth]{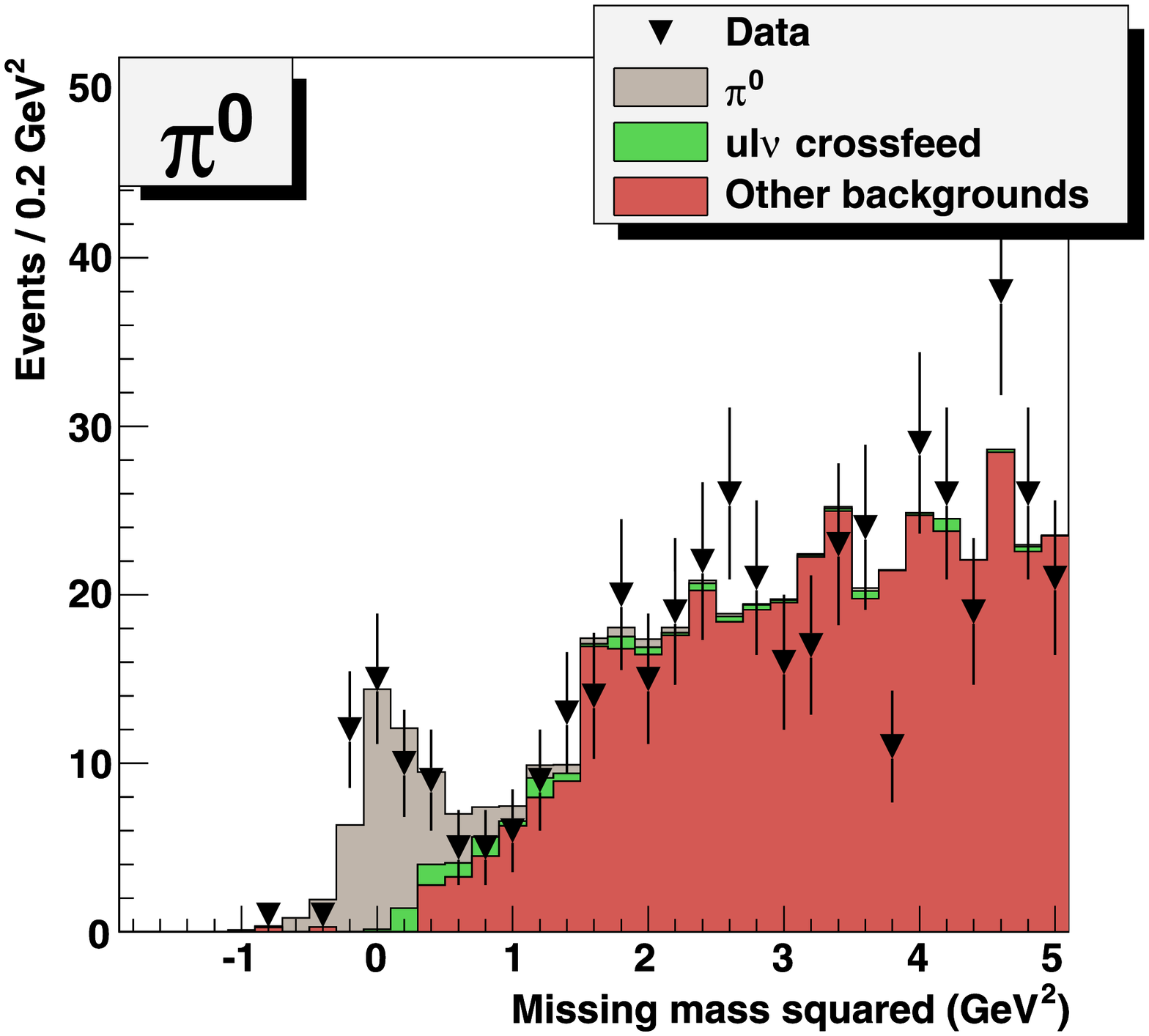}
    \includegraphics[width=0.20\columnwidth]{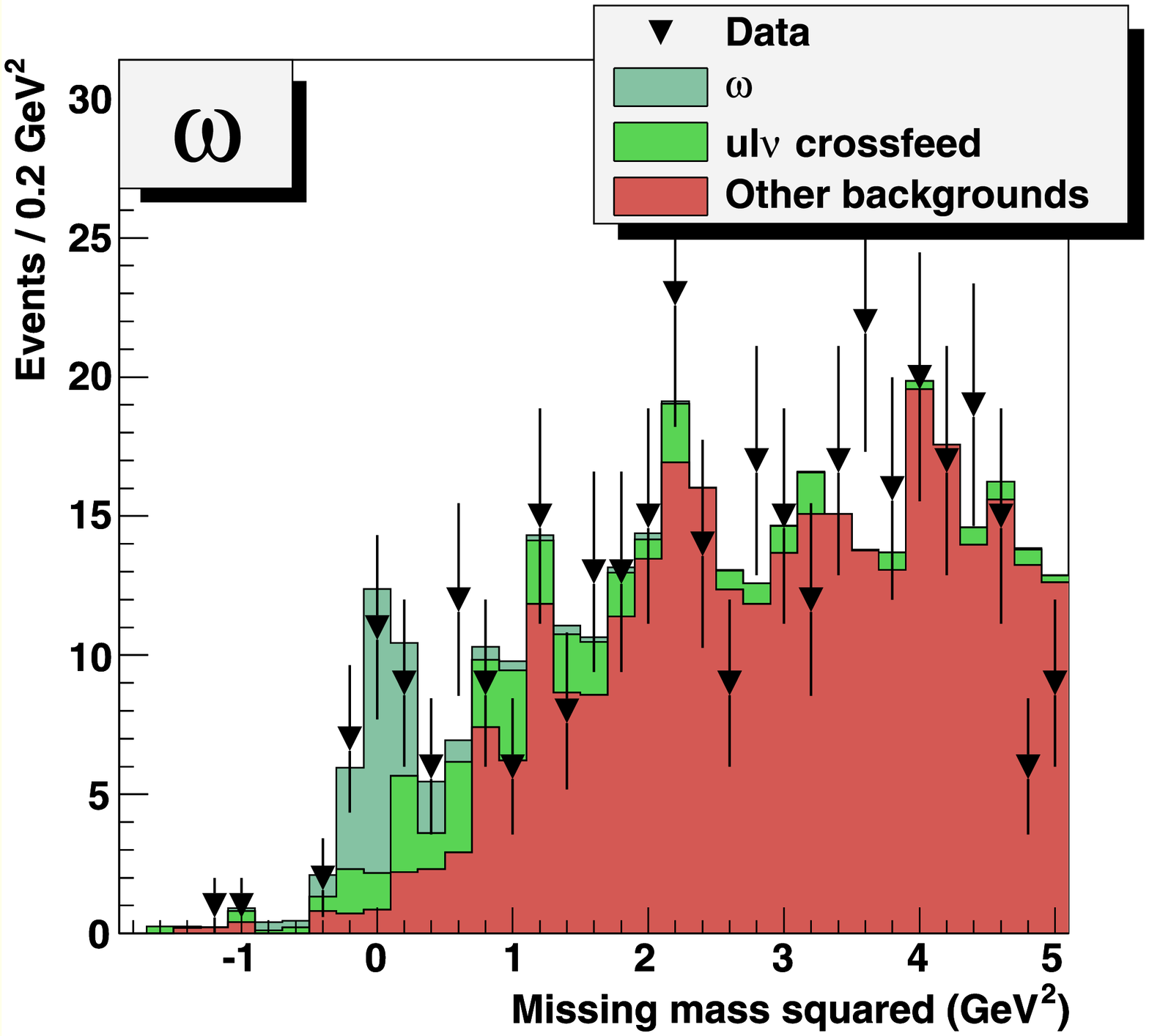}\\
    \includegraphics[width=0.20\columnwidth]{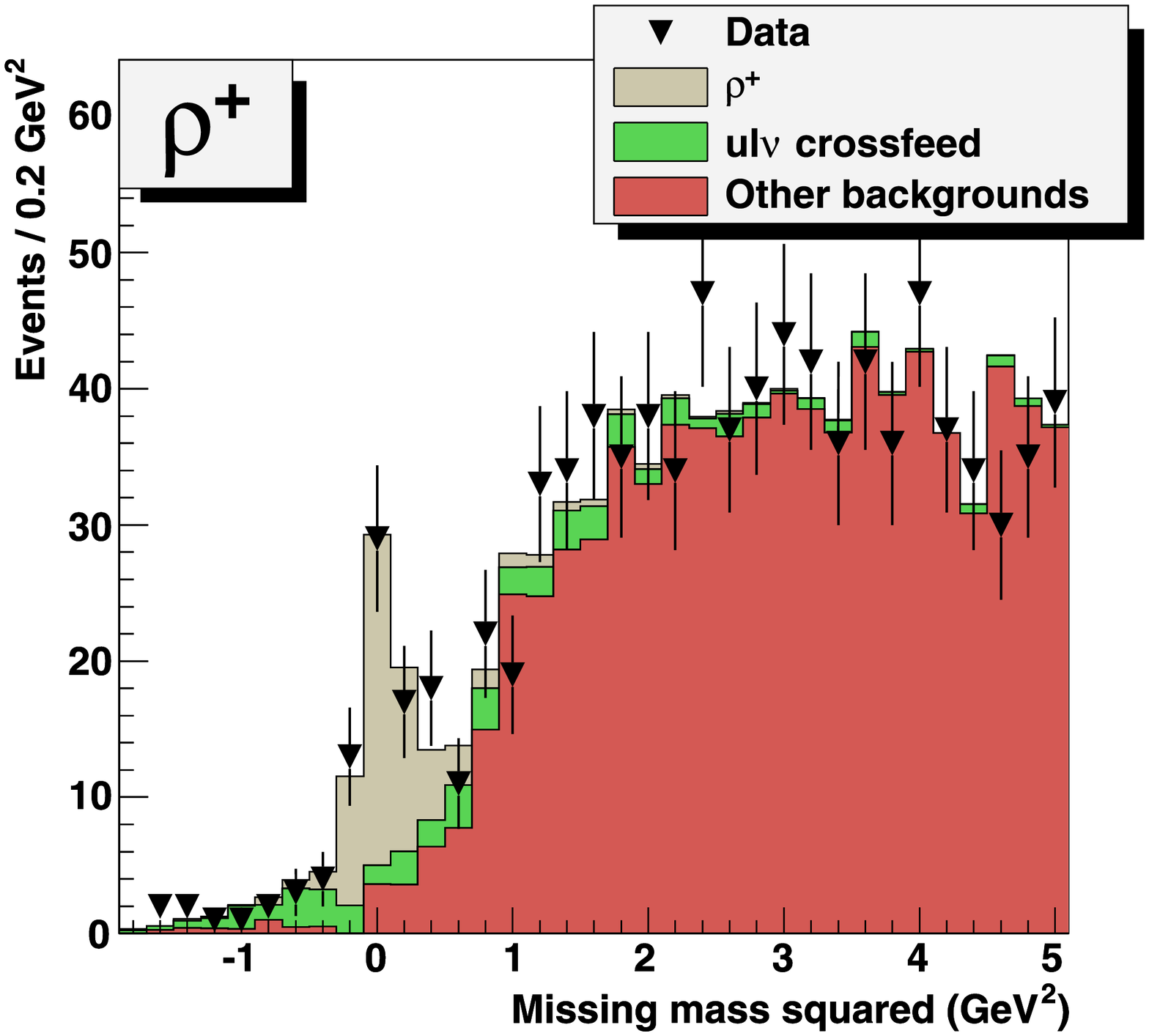}
    \includegraphics[width=0.20\columnwidth]{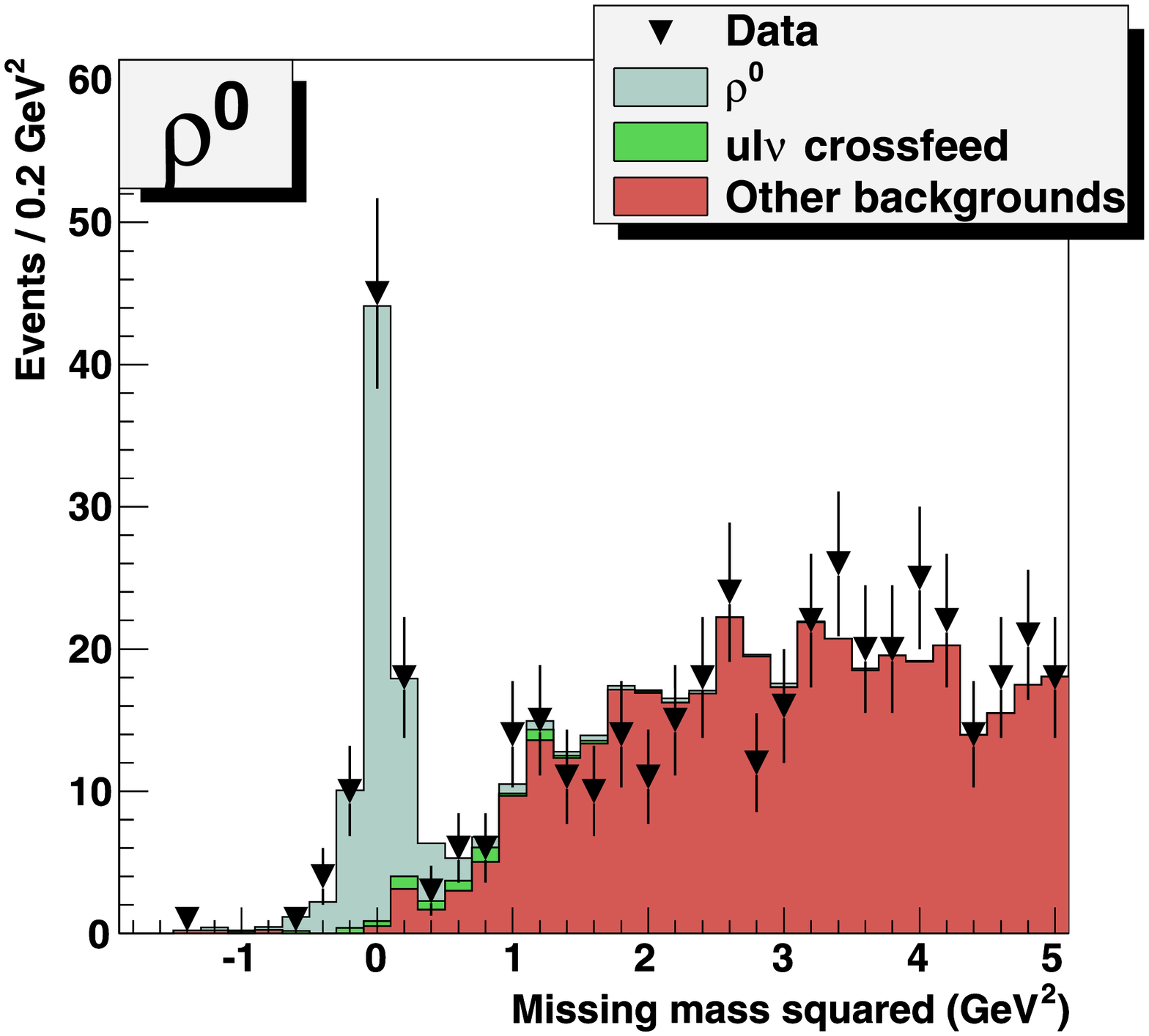}
    \phantom{ \includegraphics[width=0.20\columnwidth]{fig/mm2_omega_fit_sum.eps} }
  \end{center}
\end{figure}

For signal the missing momentum $p_{\mathrm{miss}} = p_{\mathrm{signal}} - p_{\ell} - p_{X}$ corresponds to the momentum of the neutrino. A fit to the missing mass distribution, using the ROOT (\cite{Brun:1996ro}) \texttt{TFractionFitter} class, which uses the algorithm described in Ref.~\cite{Barlow:1993dm}, is used to extract the signal yield. Three subsamples, corresponding to the phase space regions $q^2 < 8 \,\mathrm{GeV}^2$, $ 8 \,\mathrm{GeV}^2 < q^2 < 16 \,\mathrm{GeV}^2$ and $q^2 > 16 \,\mathrm{GeV}^2$, are investigated seperately, where $q^2 = (p_{\ell} + p_{\nu})^2$ is the momentum transfer. 
The results of the fit can be seen in Fig.~\ref{fig:KevinMissingMass}, where the three subsamples have been summed. The partial branching fractions, which are obtained from the fit, can be seen in Tab.~\ref{tab:KevinResults}. The error is dominated statistically.

\subsection{Exclusive $B^0 \rightarrow D^{*-} \ell^+ \nu$ decays}

\begin{figure}[hb]
  \caption{The definition of the four kinematic variables $w $, $\cos \theta_{V}$, $\cos \theta_{\ell}$ and $\chi$ and a sketch of the reconstruction of the signal $B$ momentum using momentum conservation.}
  \label{fig:MeineKinematicVariablesReconstruction}
  \begin{center}
    \includegraphics[width=0.26\columnwidth]{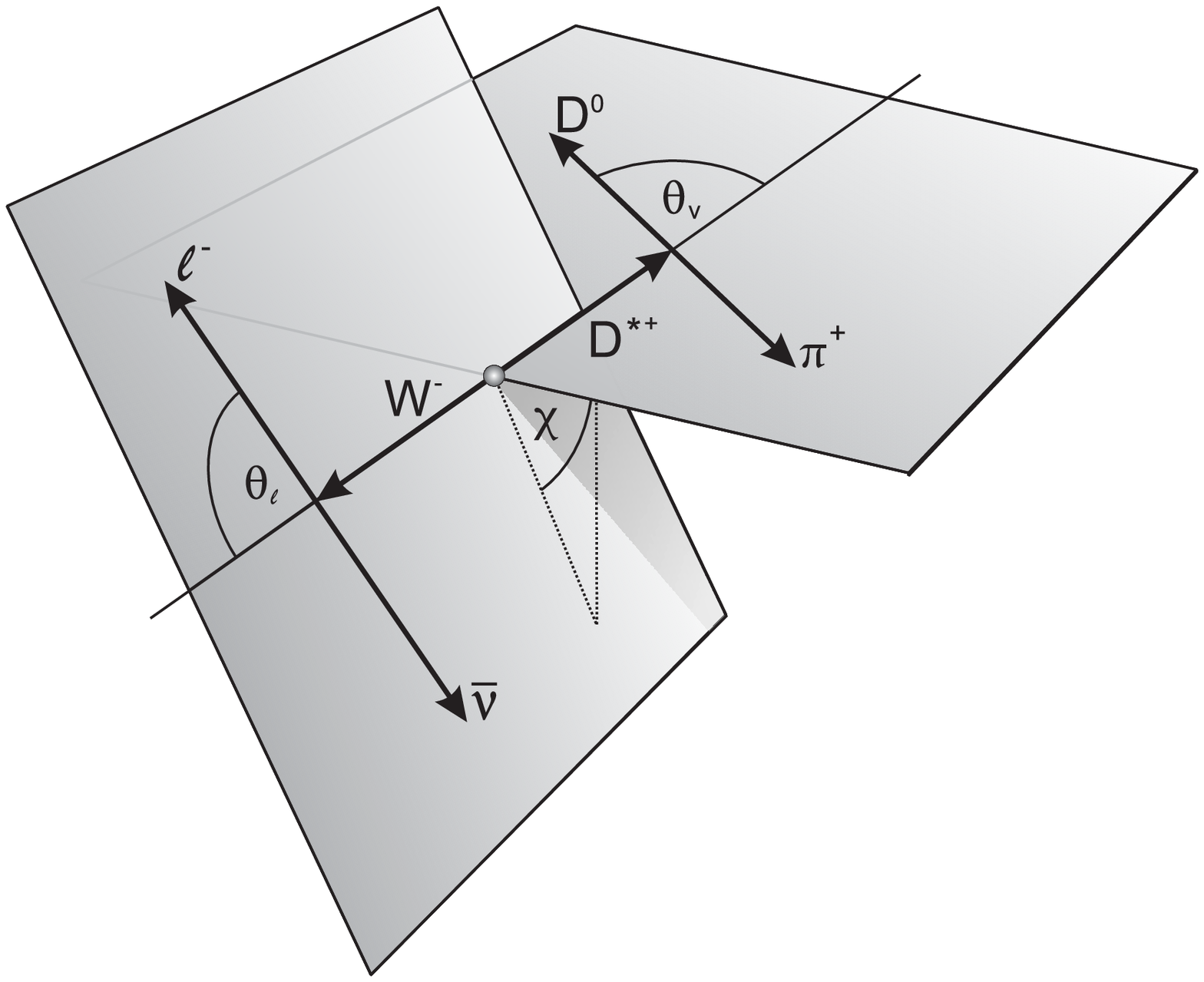}
    \includegraphics[width=0.41\columnwidth]{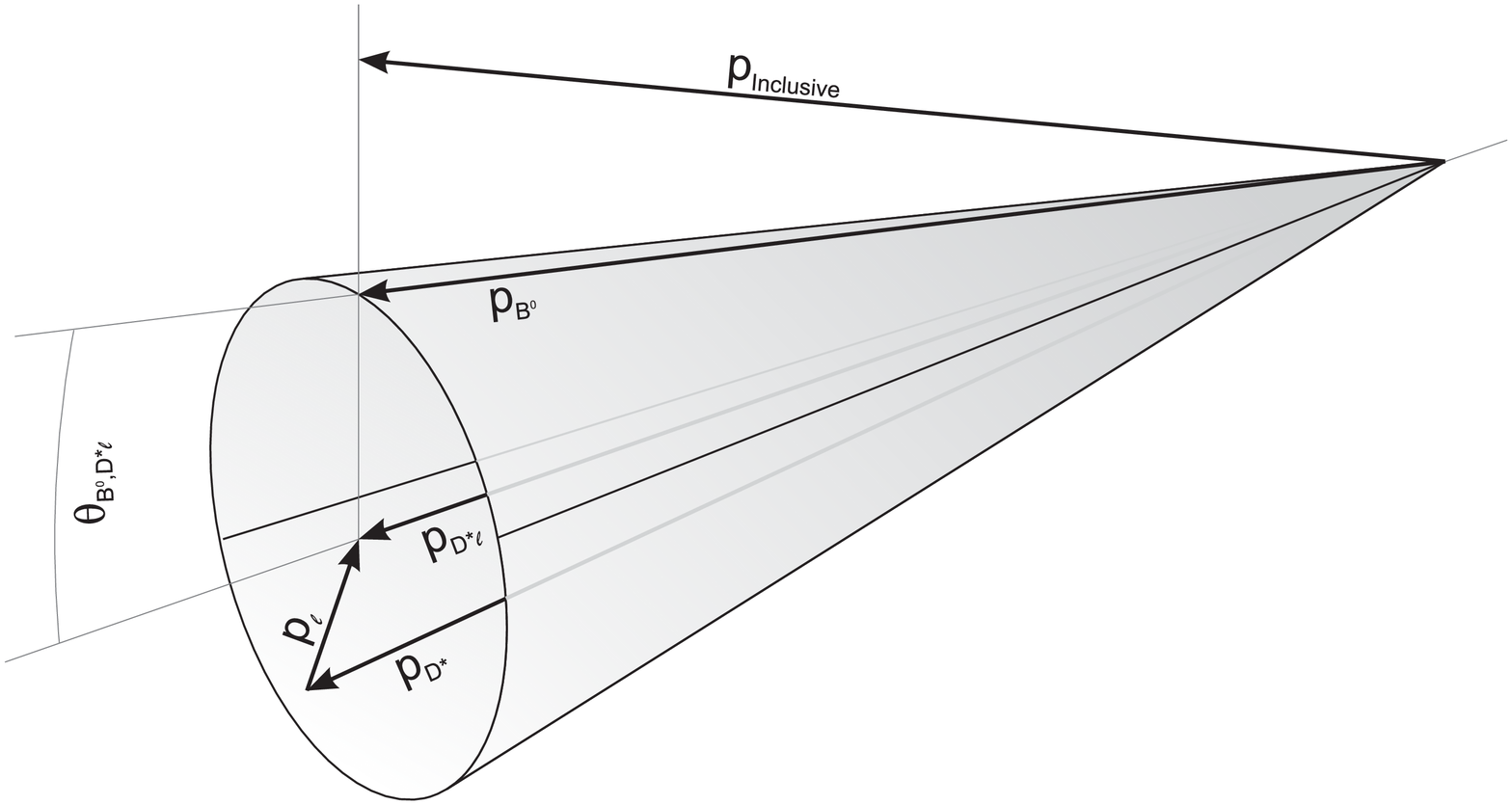}
  \end{center}
\end{figure}

The analysis aims at determining the quadruple differential decay width of the process $B^0 \rightarrow D^{*-} \ell^+ \nu$, which can be found e.g. in Ref.~\cite{Neubert:1994} or \cite{RichmanBurchat:1995}. This decay width is a four dimensional function depending on the four kinematic variables $w = v_{B}\cdot v_{D^{*}}$ and the three angles $\cos \theta_{V}$, $\cos \theta_{\ell}$ and $\chi$, which are defined in Fig.~\ref{fig:MeineKinematicVariablesReconstruction}. We use the parametrization defined in Ref.~\cite{CLN:1998}, which introduces the three free parameters $\rho^2$, $R_{1}(1)$ and $R_{2}(1)$ to govern the shape of the form factors of the decay.

The decay cascade $\bar{B}^0 \rightarrow D^{*+} \ell^- \bar{\nu}$, $D^{*+} \rightarrow D^{0} \pi^+$ and $D^{0} \rightarrow K^{-} \pi^+$ or $D^{0} \rightarrow K^{-} \pi^+ \pi^+ \pi^-$ is reconstructed. The light lepton $\ell$ is either an electron or a muon. Due to momentum conservation, the spatial momentum of the signal $B$ has to lie on a cone around the spatial momentum of the $D^{*}\ell$ system. The inclusive sum of the entire remaining event
is used to obtain the best $B$ candiate by orthogonal projection, as sketched in the right hand plot of Fig.~\ref{fig:MeineKinematicVariablesReconstruction}. Off-resonance data is used to investigate background from $q\bar{q}$ decays and MC for a set of five additional background components.

The parameters $\mathcal{F}(1) \left| V_{cb} \right|$, $\rho^2$, $R_{1}(1)$ and $R_{2}(1)$ are obtained by a binned least squares fit to the four one-dimensional marginal distributions of the decay width. The bin-to-bin correlations between these one dimensional histograms have to be considered. Only the branching ratio of the mode $D^{0} \rightarrow K^{-} \pi^+$ is used as an external parameter, the branching ratio of the mode $D^{0} \rightarrow K^{-} \pi^+ \pi^+ \pi^-$ is determined by fitting the ratio between the two $D^{0}$ channels, $R_{K3\pi/K\pi}$. A $\chi^2$ function is formed for each of the four channels seperately and the sum of these four $\chi^2$'s is minimized numerically using the MINUIT package~\cite{James:1975dr}.

\begin{figure}[h]
  \caption{Result of the fit of the four kinematic variables in the
    total sample. (The different sub-samples are added in this plot.)
    The points with error bars are continuum subtracted on-resonance
    data. The histograms are the signal and the different background
    components.}
  \label{fig:MeineResultPlot}
  \begin{center}
    \includegraphics[width=0.6\columnwidth]{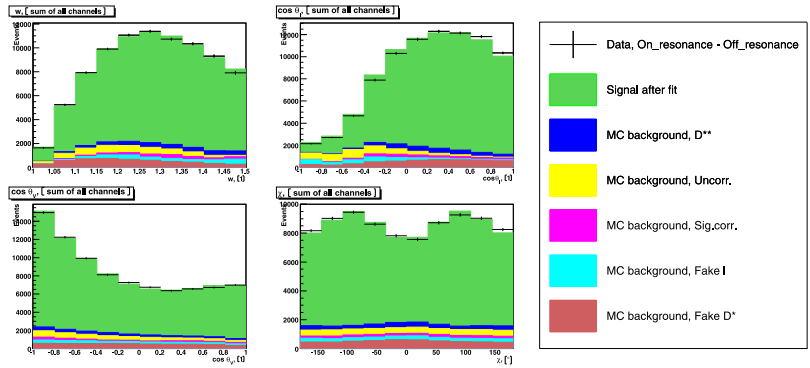}
  \end{center}
\end{figure}

The preliminary results of the fit are $\rho^{2}=1.293 \pm 0.045 \pm 0.029$, $R_{1}(1)= 1.495 \pm 0.050 \pm 0.062$, $R_{2}(1)= 0.844 \pm 0.034 \pm 0.019$, $R_{K3\pi/K\pi} =2.153 \pm 0.011$, $\mathcal{B}( B^{0} \rightarrow D^{*-} \ell^{+} \nu_{\ell} )= (4.42 \pm 0.03 \pm 0.25) \%$ and $\mathcal{F}(1) \left| V_{cb} \right| = (34.4 \pm 0.2 \pm 1.0) \times 10^{-3}$, where the first error is the statistical error reported by MINUIT and the second (where shown) is the preliminary systematic error. The $\chi^{2} / \mathrm{n.d.f.}$ of the fit gives $138.8/155$. The systematic error of the HQET parameters is dominated by background uncertainty, for $V_{cb}$ the error of the track reconstruction efficiency is dominating.
%
The results of the fit are shown graphically in Fig.~\ref{fig:MeineResultPlot}.
The obtained value for $\mathcal{F}(1) \left| V_{cb} \right|$ is one of the lowest obtained by recent measurements, but in acceptable agreement with measurements by BaBar~\cite{Aubert:2007rs}.

%


\subsection{Study of $B\rightarrow D^{**}\ell \nu$ with full reconstruction tagging}

According to HQET there are two doublets of orbitally excited charmed mesons, labeled $D_{0}^*$ $( 0^+,\,j_q = 1/2)$,
$D_{1}^{'}$ $( 1^+,\,j_q = 1/2)$, $D_{1}$ $( 1^+,\,j_q = 3/2)$ and $D_{2}^*$ $( 2^+,\,j_q = 3/2)$. The entire set of these four states is labeled $D^{**}$. These states decay into either $D\pi$ ($D_0^*$, $D_2^*$) or $D^*\pi$ ($D_1$, $D_1^{'}$, $D_2^*$). This analysis investigates $B \rightarrow D^{(*)} \pi \ell \nu$ decays, for both charged and neutral $B$ mesons, and measures the excited $D$ contributions to the $D^{(*)}\pi$ final state.

The signal $B$ meson is reconstructed in the semileptonic mode of interest ($p_{\mathrm{sl}}$) and the remaining event is then combined into either a $D^{(*)}n\pi^\pm$, $n<6$, or a $D^{(*)}\rho^\pm$ combination ($p_{\mathrm{tag}}$). The missing mass squared spectrum, $m_{\nu}^2 = (p_{\mathrm{beams}} - p_{\mathrm{sl}} - p_{\mathrm{tag}})^2$, where $p_{\mathrm{beams}}$ is the total four-momentum of the beams, is used to identify signal events. The signal window is defined as $|m_{\nu}^2| < 0.1 \,\mathrm{GeV}^2$. Background is estimated using MC simulation.

\begin{figure}[ht]
  \caption{Hadronic invariant mass distributions for 
  $B^+ \rightarrow D^- \pi^+ \ell^+ \nu$, 
  $B^+ \rightarrow D^{*-} \pi^+ \ell^+ \nu$,
  $B^0 \rightarrow \bar{D}^{0} \pi^+ \ell^+ \nu$ and
  $B^0 \rightarrow \bar{D}^{*0} \pi^+ \ell^+ \nu$. Insets show the distributions before background subtraction.
  The hatched histograms show the background.
   }
  \label{fig:DssMassDistributions}
  \begin{center}
    \includegraphics[width=0.6\columnwidth]{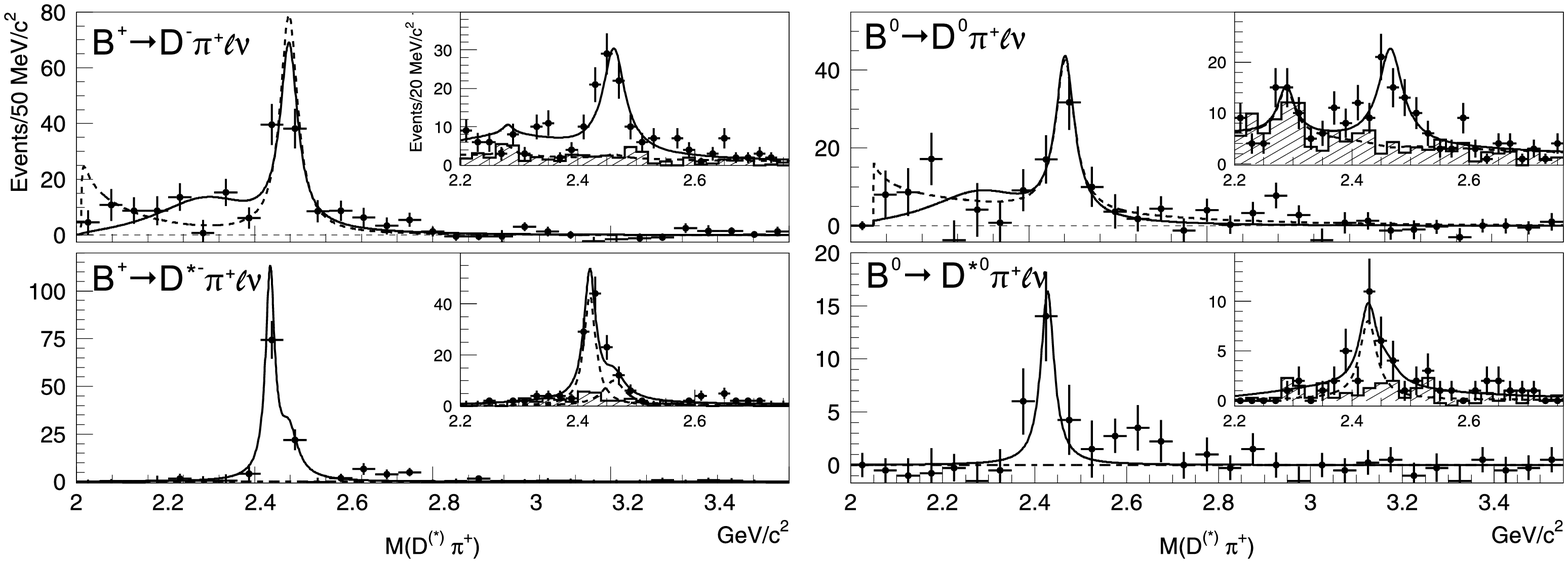}
  \end{center}
\end{figure}

The $D^{**}$ signals are extracted via simultaneous unbinned likelihood fits to the signal and background $D^{(*)}$ mass spectra. The results of the fit can be seen in Fig.~\ref{fig:DssMassDistributions}. The signal functions, shown as the solid lines in the figure, describe each $D^{**}$ state by a relativistic Breit-wigner function. A non-resonant part is modeled by the Goity-Roberts model~\cite{GoityRoberts:1995}, its contribution is consistent with zero in all cases. To further investigate the $D\pi$ mass spectrum, also a $D_{\nu}^{*} + D_{2}^{*}$ hypothesis is tested. The $D_{\nu}^{*}$ contribution is described by a tail of the Breit-wigner function. Fit results for this combination are shown as a dashed line.

The signal yields and branching fractions obtained from this fit can be seen in Tab.~\ref{tab:DssResults}.

\begin{table}[ht]
\begin{center}
\caption{Results of the $D^{(*)}\pi$ pair invariant mass study. $\mathcal{B}(\mathrm{mode}) \equiv \mathcal{B}( B\to D^{**}\ell \nu)\times \mathcal{B}(D^{**} \to D^{(*)}\pi)$. The first error is statistical and the second is systematic.}
		\label{tab:DssResults}
		\begin{tabular}[t]{|l|c|c|}
			\hline 
			\multicolumn{3}{|c|}{$B \rightarrow D^* \pi \ell \nu$ states}\\
			\hline 
			Mode & Yield & $\mathcal{B},\,[\%]$
			\\ \hline
			  $B^{+} \rightarrow \bar{D}_{1}^{'0} \ell^+ \nu$ & $-5 \pm 11$ & $ < 0.07 \, @ \, 90\% \, \mathrm{C.L.}$
			\\$B^{+} \rightarrow \bar{D}_{1}^{0} \ell^+ \nu$ & $81 \pm 13$ & $0.42 \pm 0.07 \pm 0.07$
			\\$B^{+} \rightarrow \bar{D}_{2}^{0} \ell^+ \nu$ & $35 \pm 11$ & $0.18 \pm 0.06 \pm 0.03$
			\\$B^{0} \rightarrow {D}_{1}^{'-} \ell^+ \nu$ & $4 \pm 8$ & $ < 0.5 \, @ \, 90\% \, \mathrm{C.L.}$
			\\$B^{0} \rightarrow {D}_{1}^{-} \ell^+ \nu$ & $20 \pm 7$ & $0.54 \pm 0.19 \pm 0.09$
			\\$B^{0} \rightarrow {D}_{2}^{*-} \ell^+ \nu$ & $1 \pm 6$ & $< 0.3 \, @ \, 90\% \, \mathrm{C.L.}$
			\\ \hline 
		\end{tabular}
		\hspace{2mm}
		\begin{tabular}[t]{|l|c|c|}
			\hline 
			\multicolumn{3}{|c|}{$B \rightarrow D \pi \ell \nu$ states}\\
			\hline 
			Mode & Yield & $\mathcal{B},\,[\%]$
			\\ \hline
			  $B^{+} \rightarrow \bar{D}_{0}^{*0} \ell^+ \nu$ & $102 \pm 19$ & $0.24 \pm 0.04 \pm 0.06$
			\\$B^{+} \rightarrow \bar{D}_{2}^{*0} \ell^+ \nu$ & $94 \pm 13$ & $0.22 \pm 0.03 \pm 0.04$
			\\$B^{0} \rightarrow {D}_{0}^{*-} \ell^+ \nu$ & $61 \pm 22$ & $0.20 \pm 0.07 \pm 0.05$
			\\$B^{0} \rightarrow {D}_{2}^{*-} \ell^+ \nu$ & $68 \pm 13$ & $0.22 \pm 0.04 \pm 0.04$
			\\ \hline 
		\end{tabular}
	\end{center}
\end{table}
\section{Conclusions}

We presented three analyses based on data samples collected by the Belle detector at the KEKB $e^+e^-$ asymmetric energy collider: the measurement of the branching fractions and $q^2$ distributions of exclusive charmless semileptonic B decays, the determination of $\mathcal{F}(1) \left| V_{cb} \right|$ and the form factor parameters $\rho^2$, $R_1$ and $R_2$ in the exclusive decay $B^0 \rightarrow D^{*-} \ell^+ \nu$ and the investigation of $B \rightarrow D^{**} \ell \nu$ decays.

We thank the KEKB group for the excellent operation of the accelerator, the KEK
cryogenics group for the efficient operation of the solenoid, and the KEK computer group
and the National Institute of Informatics for valuable computing and Super-SINET network
support. 

\end{document}